# Thermally Condensing Photons into a Coherently Split State of Light

Christian Kurtscheid[1,*], David Dung[1], Erik Busley[1], Frank Vewinger[1], Achim Rosch[2], and Martin Weitz[1,*]

[1]*Institut für Angewandte Physik, Universität Bonn, Wegelerstr. 8, 53115 Bonn, Germany*

[2]*Institut für Theoretische Physik, Universität zu Köln, Zülpicher Str. 77, 50937 Cologne, Germany*

[*]Corresponding authors. E-mail: kurtscheid@iap.uni-bonn.de (C. K.); martin.weitz@uni-bonn.de (M. W.)

Techniques to control the quantum state of light play a crucial role in a wide range of fields, from quantum information science to precision measurements. While for electrons in solid state materials complex quantum states can be created by mere cooling, in the field of optics manipulation and control currently builds on non-thermodynamic methods. Using an optical dye microcavity, we have split photon wavepackets by thermalization within a potential with two minima subject to tunnel coupling. Even at room temperature, photons condense into a quantum-coherent bifurcated ground state. Fringe signals upon recombination show the relative coherence between the two wells, demonstrating a working interferometer with the non-unitary thermodynamic beamsplitter. This energetically driven optical state preparation opens up an avenue for exploring novel correlated and entangled optical manybody states.

Achieving order by populating the ground state of composite systems, as in the formation of the low-energy 'bonding' superposition state of the $H_2^+$-molecular ion or the singlet ground state of atomic helium, is the experimental basis of many quantum phenomena, including the emergence of strongly correlated electron physics upon the refrigeration of solids (1). The observation of phenomena like superconductivity or the (fractional) quantum Hall effect also depends on this approach. Cooling to the ground state is the prime example for an *irreversible* temporal evolution which leads to a unique quantum state. Quite contrary, for light *reversible* unitary processes, as in a beam splitter, are the standard manipulation tools, with irreversibility only possible when introducing photon loss (2). For classical physics phenomena, loss can be balanced by gain, and laser-like non-equilibrium processes efficiently achieve macroscopic population of excited state modes (3-5). Entangled few-body states of light can be created by parametric down-conversion, which requires careful postselection when aiming at more than two entangled particles (6). Advances towards the ordering of complex optical quantum systems include the observation of exciton-polariton and photon Bose-Einstein condensates (7,8). While not thermalizing into the ground state of the multimode problem, recent work has demonstrated periodic potentials for both polariton and photon microcavity systems (9-12) and reported simulations of the classical XY-model (13,14). The fast relevant timescales, which are in the picosecond regime for both photons and polaritons, make it impractical to temporarily alternate between cooling and trap manipulation phases for optical quantum gases. While this technique is commonly used in the study of cold atoms in lattices, the required adiabatic mapping of the condensate state into the lattice is a main bottleneck to reach ultralow entropy states in those systems (15).

Here we thermalize light in a potential with two minima. By contact to an external reservoir, i.e. by an irreversible process, the system reaches a coherently split state. To achieve this, a double-well superimposed by a shallow harmonic trapping potential is tailored by shaping mirror surfaces in an optical dye microcavity. Thermalization of photons, by contact to the optically active dye molecules, reduces the mean photon energy and is found to efficiently operate in the combined trap. The symmetric linear combination of the eigenfunctions localized in the two sites is the energetic ground state of cavity photons, and after reaching the condensation threshold, we observe a macroscopic occupation of photons in this "bonding" eigenstate at room temperature. The fixed phase relation between the optical wavefunction components in the different sites is verified by optical interferometry.

Our experiment, see Fig.1A for a schematic picture, traps photons in a dye-filled microcavity, Two-dimensional photon gases in such dye microcavities show a thermodynamic phase transition to a macroscopically occupied ground state, the Bose-Einstein condensate (8,16,17). To engineer the desired potential, one of the two ultrahigh reflectivity cavity mirrors (cavity finesse near 100000) has a microstructured surface profile, manufactured with a novel delamination technique (18). In brief, the profile is created by scanning an auxiliary laser beam transversally over the mirror plane, inducing heat from absorption in a 30nm thick silicon layer placed below the dielectric coating (12). The heating causes a controlled local delamination of the reflective surface, which for the cavity results in a shorter distance between mirrors at the corresponding transverse position. In the paraxial limit this induces a repulsive potential for light, which intuitively can be understood from that here shorter wavelength (higher energetic) photons are required to match the boundary conditions of the reflecting surfaces.

A spatial profile of the structured mirror profile used for this work is shown in Fig.1B, together with a cut and the calculated potential for photons within the microcavity on the right hand side. Near the mirror center, two indents spaced by 13μm cause a double-well potential sufficiently shallow that only a single eigenfunction per site is trapped. Tunneling between the sites leads to a coupling of the localized wavefunctions, which hybridize to a symmetric $\psi_s = \frac{1}{\sqrt{2}}(\psi_1 + \psi_2)$ and an antisymmetric wavefunction $\psi_a = \frac{1}{\sqrt{2}}(\psi_1 - \psi_2)$, where $\psi_1$ and $\psi_2$ denote the wavefunctions localized in the wells, see also Fig.1C. The eigenenergies of the hybridized wavefunctions are split by $\hbar\Delta$, where $\Delta \cong 2\pi \cdot 30$GHz denotes the tunnel coupling, with the "bonding" state $\psi_s$ being energetically lower than $\psi_a$. As to allow for the existence of a Bose-Einstein condensate in two dimensions, the double-well potential is superimposed by a weak harmonic trapping potential by imprinting a parabolic mirror profile with ≅15cm radius of curvature (18).

Thermalization of photons is achieved following methods described previously (8,19). The two cavity mirrors, due to their small spacing in the micrometer regime, impose an upper limit of the optical wavelength that fits inside the resonator, corresponding to a restriction of energies to a minimum cut-off of $\hbar\omega_c \cong 2.1$eV, which we identify here with the eigenenergy of the "bonding" state $\psi_s$ of the double-well. Moreover, the optical dispersion of cavity photons in the paraxial limit becomes quadratic, as for a massive particle. Given that interparticle interactions are weak, (number-conserving) thermalization of photons is achieved by repeated absorption re-emission processes on the dye molecules, which occurs to the rovibrational temperature of the dye, which is at room temperature. Due to the frequent collisions of solvent molecules with the dye (18,19), photons emitted by the dye molecules are thermally equilibrated in an irreversible process. Due to the large mode spacing in the wavelength-spaced cavity, of order of the emission width of the dye, the thermalization leaves the longitudinal mode unchanged, while the remaining transverse degrees of freedom make the photon gas two-dimensional. It has been shown (19) that the system is formally equivalent to a two-dimensional system of trapped massive bosons with effective mass $m_{ph}=\hbar\omega_c n^2/c^2$. To inject photons and compensate for losses, the dye is weakly pumped with a laser beam. Despite pumping and losses the photon gas well follows an equilibrium Bose-Einstein distribution

$$n_B(u_i) = \frac{g(u_i)}{\exp\left(\frac{u_i - \mu}{k_B T}\right) - 1}, \qquad (1)$$

because photons thermalize faster than they are lost through e.g. mirror losses (20,21). Here $u_i$ denotes the excitation energy above the cut-off, $g_i$ the mode degeneracy, and $\mu$ the chemical potential. For the lowest two levels of our system, the "bonding" and "anti-bonding" states of the double-well, we have $u_s=0$ and $u_a=\hbar\Delta$, with $g_i=2$ in both cases due to polarization degeneracy, see (18) for details of our theoretical modelling.

Fig.2A shows experimental data analyzing the microcavity emission. The top and middle graph give color-coded high-resolution spectra spatially resolved along the direction of the double-well axis. The top panel shows data recorded in the thermal regime with a photon number far below the Bose-Einstein condensation threshold. On the long wavelength (low photon energy) side, one observes the emission of the antisymmetric (slightly below 586nm) and symmetric (slightly above 586nm) double-well modes. In the former case the emission in the center vanishes, indicating photons in the antisymmetric state, while the finite intensity in the center for the latter case corresponds to a population of the symmetric wavefunction. To verify the thermal distribution of modes, we have recorded broadband spectra for different total photon numbers, as shown in Fig.2B along with theory curves (solid lines) for a Bose-Einstein distribution (eq.1). The bottom curves here correspond to the thermal regime, while at high photon numbers we observe a peak at the position of the cavity low-frequency cut-off near 586nm, attributed to photon Bose-Einstein condensation. Our experimental data is in good agreement with the equilibrium theory for 300K down to a wavelength near 582nm, at which we approach the finite rim of our imprinted potential (see also Fig.1B) at around h·3.4 THz ($\cong 0.54 \cdot k_B T$) above the cut-off energy, due to present limits of our delamination technique. Note that the thermal cloud saturates for photon numbers above the BEC threshold, verifying a central prediction of BEC theory.

The middle panel of Fig.2A displays data above the condensation threshold with the individual cavity mode energies spectrally resolved, see also the bottom panel (green data points). The visible strongly enhanced emission at the position of the cavity low-frequency cutoff is evidence for Bose-Einstein condensation of photons into the symmetric superposition $\psi_s$ of the two microsites' eigenfunctions, where the condensate contains approximately 1200 photons. Interestingly, the energetic spacing to the first excited state ($\psi_a$) is $\hbar\Delta \cong k_B \cdot 1.4K$, roughly a factor 200 below the thermal energy of the room temperature system. This comparison, in addition to the good agreement of our data with BEC theory, makes it clear that the observed population enhancement is a consequence of quantum, rather than classical, statistics.

The full spatial profile of the cavity emission is shown in Fig.3A both below (left) and above (right) the critical photon number. Corresponding cuts along the direction of the double-well are shown in Fig.3B for both the data below (blue line) and above (black line) criticality, where in the latter case the split spatial profile of the condensate is clearly visible. To determine the relative phase of the emission originating from the individual microsites, we have directed the cavity output through an optical interferometer. Figure 4A displays camera images of the resulting signal recorded in the thermal (left) and the condensed phase (right) respectively, the marked central area corresponds to the region where the emission of the individual microsites spatially overlap. In this region we observe a stable interference signal for the condensed phase, see also Fig.4B, even though the shown data are the average of many subsequent experimental realizations (captions and (18)). This verifies the fixed phase relation between the microsites, as expected for a delocalized, macroscopically occupied superposition state. Put differently, the observation of a fringe pattern sets an upper limit on the indistinguishability ("which-path information") left in the individual microsites upon thermalization. This can be understood from the tunneling time $\pi/\Delta \cong 17\text{ps}$ being below the $\cong 100\text{ps}$ dye reabsorption time during the absorption and emission cycles for the used experimental parameters (21).

To conclude, we have observed Bose-Einstein condensation of photons into the "bonding" low energy superposition state of a double-well potential. The relative phase coherence of the microsites' emission was verified by overlapping the corresponding outputs, realizing an interferometer with a quantum thermodynamic beamsplitter. This demonstrates the irreversible but coherent population of split states of light, where the bifurcated single-photon state is massively occupied due to Bose-Einstein statistics.

Thermalization of light to cold reservoirs in tailored trapping geometries holds promise for the direct preparation of more complex quantum states. Strong photon interactions are expected from using second-order nonlinear materials in a doubly-resonant cavity setup, yielding a controllable effective Kerr interaction (22). When in a periodic lattice potential quantum manybody states then become the ground state, they can be selectively populated by thermalization (18, 23-26), which is not yet achievable in present atomic physics experiments (15). For analysis, optical manybody state tomography can be performed with correlation measurements (6). Other perspectives of our work include the exploration of novel correlated quantum states in both double-well and periodic potential lattice systems coupled via effective particle exchange to the photo-excitable dye molecules (27,28). Finally, applications of the described delamination-based adaptive optics method can range from gain-dissipative simulations of the classical XY-model in partial equilibrium condensate arrays (13,14) to optical phase holography (3).

**Acknowledgements:**
We acknowledge funding from the DFG within the focused research center SFB/TR 185 (277625399), SFB/TR 183 (277101999), and the Cluster of Excellence ML4Q (EXC 2004/1 – 390534769), the EU within the ERC Advanced Grant project INPEC and the Quantum Flagship project PhoQuS, and the DLR project BESQ.

**Supplementary Text**

Experimental setup and technique for creating variable potentials

The employed optical microcavity consists of two highly reflecting mirrors (reflectivities above 99.997% in the wavelength range 532nm - 590nm) spaced $D_0 \cong 2.2\mu m$ apart, with the spacing filled with rhodamine 6G dye dissolved in ethylene glycol (1 mmol/l). For the used room temperature conditions, rapid decoherence from collisions of solvent molecules with the dye on a $10^{14}$/s timescale prevent a coupling of the phases of dipole and photon (29,30), so that the confined particles are well described as photons (i.e. not polaritons). To generate variably shaped potentials for cavity photons, we use a technique where the surface profile of one of the cavity mirrors is microstructured. Our setup is a modified version of an apparatus described in earlier work, where photon potential creation was reached by varying the intracavity refractive index using a thermosensitive polymer admixed to the dye solution (12).

We use two plane cavity mirrors with one of the mirrors possessing a 30nm silicon layer below its dielectric coating (coating manufacturer: Laseroptik GmbH, Garbsen). Microstructuring is achieved (prior to operation of the microcavity) by guiding a laser beam near 532nm wavelength focused to 1.5μm diameter to heat the absorbing silicon layer, causing a permanent lifting of the mirror reflective surface near the focus, though - in the used power range of this "heating" laser beam - seemingly not affecting the mirror reflectivity. Above a threshold power of ~30 mW we observe a local vertical deformation of the dielectric coating, with a typical transverse width of 3.6μm. The deformation is attributed to the absorbed power leading to a delamination of the dielectric coatings from the glass substrate due to heat-induced enhancement of mechanical stress of the dielectric layers. To allow for the engineering of variably shaped mirror surface profiles, the laser beam is transversally steered over the mirror surface with a two-dimensional galvo-scanner, and the beam power is controlled with an acousto-optic modulator. The delamination height is monitored in a spatially resolved way with a Mirau-interferometer. In subsequent steps, the heat-pattern is iteratively adjusted to manufacture the desired structure. In this way, a quite accurate tailoring of photon potentials is possible, with the typical longitudinal mean deviation of the final profiles from the desired mirror shape being about 1Å. Figure 1B shows the profile of the mirror used in this experimental work, as obtained from the Mirau interferometer.

The quantum efficiency of the used rhodamine 6G dye is ~95%, and the dye fulfills the Kennard-Stepanov-relation, corresponding to a Boltzmann-like thermodynamic frequency scaling between absorption and emission spectral profiles respectively (31). The background of this relation are the above mentioned several femtosecond timescale collisions of solvent molecules with the dye, which alter the rovibrational state much faster than the timescale of electronic processes (the upper state electronic lifetime of rhodamine dye is 4.1 ns). Both absorption and emission can thus well be assumed to occur from a state with the rovibrational manifold in thermal equilibrium with the solvent, from which the Kennard-Stepanov relation can be shown to follow. Due to the rapid relaxation within rovibrational manifolds the emission wavelength also is in good approximation independent from the excitation, which is one formulation of Kasha's rule (31), clearly corresponding to an irreversibility of the molecule-light interaction.

The used small spacing of cavity mirrors imposes a spectrum of the longitudinal cavity modes that is comparable to the emission width of the dye. For this situation, we observe that only photons of a fixed longitudinal mode number q, with q=11 here, populate the cavity. This effectively imposes a low-frequency cut-off at $\hbar\omega_c$, with $\omega_c=2\pi c/\lambda_c$, where $\lambda_c \cong 586$nm is the cut-off wavelength. Thermalization of the photon gas is obtained as photons are absorbed and re-emitted from the dye molecules. In this way, the thermalized distribution of the rovibrational manifolds of upper and lower electronic states of the dye molecules is transferred to the spectral distribution of the photon gas, driving the system towards a thermal equilibrium distribution of cavity modes, with the photon frequencies distributed by an amount $\sim k_B T/\hbar$ above the cut-off.

The dye microcavity is pumped at near 45° angle with a laser beam near 532nm wavelength, using an 80 μm pump beam diameter. To avoid bleaching of the dye, the pump radiation is chopped to 1μs long pump pulses at a 50 Hz repetition rate. Thermalization by thermal contact to the dye proceeds on a timescale of order of a few times the rhodamine upper electronic state natural lifetime ($\cong$4.1ns), or much faster far above the threshold (21), meaning the pump pulses are at least two orders of magnitude longer than the thermalization time. In other words, the experiment operates quasi-cw. The energy difference of pump beam photons to photons at the cavity cutoff corresponds to ~8.3$k_B \cdot T$, meaning that photons loose energy upon thermalization to reach a room temperature distribution in the microcavity.

Photon dispersion in the microcavity

The photon energy in the resonator as a function of longitudinal ($k_z$) and the two transversal ($k_x$, $k_y$) wave vector components can be written as $E = \hbar \frac{c}{n}\sqrt{k_x^2 + k_y^2 + k_z^2}$, with c as the speed of light and n≅1.44 is the dye solution refractive index. The mirrors' boundary conditions impose $k_z(x,y)=q\pi/(n \cdot D(x,y))$, where $D(x,y)=D_0-h(x,y)$ denotes the distance between mirror surfaces at a transverse position $(x,y)$, as measured relatively to the cavity axis, and $h(x,y)$ describes the local delamination. With $h(x,y) \ll D_0$, and in the paraxial limit ($k_x, k_y \ll k_z$) we obtain the effective Hamiltonian

$$H \cong m_{ph}\left(\frac{c}{n}\right)^2 - \frac{\hbar^2}{2m_{ph}}(\nabla_x^2 + \nabla_y^2) + V(x,y), \qquad (2)$$

with $m_{ph}=nq\pi\hbar/(cD_0)$, which describes a particle with a non-vanishing effective mass $m_{ph}$ in two dimensions subject to a potential

$$V(x,y) = m_{ph}\frac{c^2}{n^2 D_0}h(x,y). \qquad (3)$$

The two-dimensional, effective Schrödinger equation $i\hbar\, \partial_t \psi_\alpha(x,y,t) = H\, \psi_\alpha(x,y,t)$ describes the quantum dynamics of photons in a given transversal mode with polarization direction $\alpha = x, y$. The effective Schrödinger equation is valid for $h(x,y) \ll D_0$ under the condition that the height varies smoothly in space, $\nabla_{x,y} h(x,y) \ll 1$. In this limit, the two polarization directions do not couple and are degenerate. The corresponding stationary Schrödinger equation can be written in the general form $H\psi_{m,\alpha}(x,y) = E_m \psi_{m,\alpha}(x,y)$.

By lateral shaping of the mirror profile we can directly control the photon potential, whose magnitude scales linearly with the local delamination height. Eq.3 was used when deriving the photon potential of a double well superimposed by a harmonic trap from the interferometrically measured profile (right diagram of Fig.1B). The highest visible delamination height of 15nm translates directly to a rim height of h·3.4 THz, whereas the double-well, with a height of h·0.11 THz, is relatively shallow, as to only support a single bound state per site. Tunneling between the sites leads to a hybridization of eigenstates with a measured splitting of $\Delta \cong 2\pi \cdot 30\text{GHz}$ between the symmetric and the antisymmetric state respectively.

Spectral and spatial analyses of the cavity emission

The experimental spectra shown in Figs.2A and 2B were recorded by directing the transmission of the plane cavity mirror through a slitless grating spectrometer. To record the highly resolved spectrum of Fig.2A, the spectrometer was equipped with an Echelle grating, while for the lower-resolved, broadband spectrum of Fig.2B a holographic grating was used. Following the spectrometer, the emission was guided over a telescope with one of its lenses being replaced by two orthogonally oriented cylindrical lenses differing in their focal length, and then imaged onto an ICCD camera. Effectively, the x-axis in Fig. 2A (top and middle) thus contains information of mode eigenfrequency wavelength $\lambda_m$ and spatial coordinate x, following $\lambda = \lambda_m + \kappa \cdot x$, with $\kappa \cong 1.32 \cdot 10^{-3}$ nm/µm. This allows to monitor the spatial distribution of the photon gas along the vertical (y) direction as a function of the emission wavelength, with a spectral resolution that is limited by the size of the corresponding modes along the x-direction. The imaging size along this axis is reduced by the 1:5 aspect ratio of the cylindrical lens telescope. We attribute the observable spectral resolution for the case of the spatially along the x-direction well localized modes $\psi_s$ and $\psi_a$ (see also Fig.3B for the former ones full spatial intensity profile) to be limited by the instrumental ~15GHz resolution of the Echelle spectrometer, while for the higher modes in the harmonically trapped region to be dominated by the corresponding cloud size along the x-direction. The spectra shown in the bottom panel of Fig.2A were obtained by binning data as shown in the top and middle panels over the signals corresponding to the different experimentally resolved mode signals for the quoted total photon numbers.

The data shown in Fig.2A is the average of the signal recorded within 200 pump pulses, with the spectral signal corrected to compensate for shot-to-shot fluctuations of the cavity cut-off attributed to mechanical mirror vibrations. For the low resolution, broadband spectrum displayed in Fig.2B, the camera signal was binned along the horizontal camera axis. Here direct averages of the signal recorded within 50 pump pulses are shown, with the broadband nature making a correction for the wavelength dependent transmission of cavity mirrors necessary.

For a direct comparison of theory and experiment, we numerically determine the solution of the stationary Schrödinger equation (see the above section) using the measured height profile h(x,y) of the microstructured cavity mirror. The eigenstates are obtained by calculating the matrix elements of $H$ for harmonic-oscillator basis states, followed by an exact diagonalization of the matrix. Using that the intensity of the light emitted by our cavity can well be assumed to be proportional to the photon density in the cavity, we plot in Fig. S1 for two values of the chemical potential

$$I(x,y) = \sum_{n_x,n_y,\alpha} \left| \psi_{n_x,n_y,\alpha}\left(\frac{\lambda - \lambda_{n_x,n_y}}{\kappa}, y\right) \right|^2 n_B\left(u_{n_x,n_y}\right), \qquad (4)$$

Here $n_B\left(u_{n_x,n_y}\right)$ is the Bose-Einstein distribution function and $\lambda_{n_x,n_y}$ is the vacuum wavelength of photons with energy $u_{n_x,n_y}$ above the cutoff. The labeling of wavefunctions and eigenenergies by the quantum numbers $n_x$ and $n_y$ is here chosen to indicate to which harmonic-oscillator wavefunction the eigenfunctions have the largest overlap. In the upper panel of Fig. S1 we show the wavelength $\lambda_{n_x,n_y}$ of each eigenstate as a dashed line. The quantum numbers $n_x$ and $n_y$ can approximately be identified with the number of nodes in the x- and y- directions, respectively.

In the theory prediction we have not accounted for spectral broadening from the finite wavelength resolution present in the experimental spectrometer, as to maximize the visibility of features arising from the shape of the eigenstates. A comparison of Fig. S1 and Fig. 2A shows that the theory does not only reproduce very well the two low-lying states (with $n_x = n_y = 0$ for $\psi_s$ and $n_x = 0$, $n_y = 1$ for $\psi_a$ respectively), but also the main features of the intensity distribution of higher-energy states. The main discrepancy is that the measured splitting of the two lowest lying levels is about 20% smaller than the calculated one, while the energies of the modes at higher energy are captured with substantially higher precision. We attribute this to both corrections to the paraxial approximation not captured by the effective Schrödinger approximation approach becoming relevant for the few wavelength-sized double-well structure, as well as to instrumental limits of the Mirau imaging system used to characterize the mirror surface. The theory results (see Fig. S1) show that the additional double well structure also for the modes energetically above $\psi_s$ and $\psi_a$ adds corrections to the mode spectrum of a 2D harmonic oscillator, which are quite relevant for the next few states. The experimental spectra well resolve the energetic spacing between adjacent harmonic oscillator modes, but not clearly the corrections of the harmonic oscillator modes from the double-well potential. From comparison with theory, we find that the third and fourth peaks from right, best visible in the middle panel of Fig.2A, contain both the ($n_x$=1, $n_y$=0) and ($n_x$=0, $n_y$=2) modes in the former case and the ($n_x$=1, $n_y$=1) and ($n_x$=2, $n_y$=0) modes for the latter. Starting with the fifth peak, to which modes with $n_x + n_y$ =3 contribute, the mode assignment within the experimentally resolved peaks well follows a 2D harmonic oscillator model. From the spacing between the five leftmost modes shown in the middle panels of Fig.2A, we determine the harmonic trap frequency to $\Omega/2\pi$= 63(2) GHz.

For a comparison of the experimental spectra shown in the bottom panel of Fig.2A and Fig.2B with theory again a Bose-Einstein distribution of modes above the cutoff (eq. 1) was assumed, with degeneracies $g_s = g_a$ =2 (accounting for polarization) for the two double-well modes $\psi_s$ and $\psi_a$. Following the above described identification of the visible peaks in our experimental high resolution data of Fig.2a (top and middle) the third and fourth data point from the right hand side (i.e. those directly energetically above the double-well modes) in both cases is attributed to stem from populations in two transverse cavity modes, so that we arrive at a four-fold degeneracy. When labelling the resolved states energetically above two double-well modes as e1, e2, e3,…, in this model we thus have $g_{e1} = g_{e2}$ =4, and the usual expression $g_{ei}$ =2(i+1) corresponding to the degeneracy of a 2D harmonic oscillator for the higher order modes with i≥3. With these degeneracies, the total number of modes is unchanged with respect to a simple harmonic oscillator potential. For the energies of modes in Fig. 2A, bottom, we have used the centroids of the experimentally determined peaks. The chemical potential $\mu$ of the photons is implicitly defined from $N=\Sigma_i n(u_i)$, where $u_i$ denoted the corresponding mode energy above the cut-off, and $N$ is the total photon number. The theory spectra derived using this model well fit the measured data.

In a purely harmonic trap the existence of a Bose-Einstein condensate for the here relevant two-dimensional case of particles with non-vanishing effective mass is well established (32,33). For sake of clarity, we remark that this is in contrast to the three-dimensional case of blackbody radiation, for which photons at low temperature vanish instead of macroscopically population the ground state. Let us first consider the modification of the potential for microcavity photons from the (2D) harmonically trapped case due to the additional shallow double-well potential in the center (i.e. not yet accounting for the finite height of the experimental potential). In the limit of our experimental accuracy, we can for i≥3 well assume that $u_{i+1} - u_i = \hbar\Omega$. When expressing the total particle number as the sum of the populations in the four lowest energetic modes $\psi_s$, $\psi_a$, e1, and e2, followed by the harmonic oscillator modes with i≥3, we find that the contribution of all modes except for the ground mode can in the limit of $\Delta << k_B T/\hbar$ and also all spacings between adjacent higher modes $(u_{i+1} - u_i) << k_B T/\hbar$ be expressed as an integral, with these thermal modes saturating at a critical particle number of

$$N_{c,\infty} = \frac{\pi^2}{3}\left(\frac{k_B T}{\hbar\Omega}\right)^2, \tag{5}$$

at the point when $\mu \to 0^-$. This result is identical as in the absence of the shallow double-well at the trap bottom, as expected for $\Delta, (u_{i+1} - u_i) << k_B T/\hbar$. To investigate in detail the equilibrium population of low energy modes, we approximate the Bose-Einstein distribution (eq.1) for $(u_i-\mu) << k_B T$, which yields

$$n_i = \left(\frac{k_B T}{u_i - \mu}\right)\cdot g_i. \tag{6}$$

For the two double-well modes, we have $n_s \equiv N_0 = -2k_B T/\mu$ and $n_a = 2k_B T/(\hbar\Delta-\mu)$. The former expression allows us to directly relate the chemical potential to the ground mode population, yielding $\mu = -2k_B T/N_0$. With this, the (upper) antisymmetric double-well level population can be expressed as $n_a = ((\hbar\Delta/2k_B T)+1/N_0)^{-1}$, which for large ground mode population saturates at $2k_B T/\hbar\Delta$, a level of population that for a splitting $\Delta$ of order of the trap frequency $\Omega$ is quite comparable to the ones of the low energy harmonic oscillator modes (i.e. for small values of i) of $2(i+1)k_B T/(\hbar u_i)$. The latter formula also follows from eq. 6 for $N_0\to\infty$. This confirms that it is thus well appropriate to account the higher energetic double-well level population to the thermal photon cloud, with the modification of the critical photon number due to the additional double-well being far below our experimental accuracy. From eq. 5 with the above quoted value for the spacing between transverse modes in the harmonically trapped region we obtain $N_{c,\infty} \cong 33000$.

The theory lines shown in Figs.2A (bottom panel) and 2B are based on the Bose-Einstein distribution (eq. 1) for T=300K using mode spacing data measured in the low energy region with the highly resolved Echelle spectrometer (see above), i.e. assuming a 2D harmonic trap of infinite height overlaid with the double well. The curves were fitted to the experimental data (restricted to the shown spectral region below the rim), leaving the chemical potential as a free fit parameter. For the case of Fig.2B, the distribution of eq. 1 in addition was convoluted with a Gaussian curve of spectral width 0.17 nm, corresponding to the spectrometer resolution for the condensate mode, after which the fitting procedure was applied. Given the broad spectral width of the thermal photon cloud, this procedure is relevant only near the condensate mode. The experimental spectra well agree with the theory curves up to near the spectral position where the finite rim height of the generated potential is approached, see the spectral position of the vertical dashed line. This is in agreement with the usual expectation that for the formation of a Bose-Einstein condensate mainly the spectral density of states of the low energetic modes is relevant. Given the correspondingly reduced total photon number we expect that Bose-Einstein condensation sets in at a lower photon number $N_{c,\eta} = N_{c,\infty} \cdot \eta$, with the factor $\eta \leq 1$ accounting for the visible reduction in the higher order modes population near and above the rim. By comparing the spectral areas of the experimental curves to the expected ones for an infinitely high harmonic trap region in the high photon energy (low wavelength) region, we can estimate $\eta \cong 0.29(5)$, yielding a critical photon number $N_{c,\eta} \cong 9500(1600)$ which within experimental uncertainties is in agreement with the experimentally observed value $N_{c,exp} \cong 8000(2500)$, as derived from the cavity emission. A remaining imperfect saturation of the thermal modes can be attributed to an interaction-induced deformation of the trapping potential, with the main cause of effective photon-photon interactions here being of thermo-optics nature (12,34). Experimental values for the photon numbers in the symmetric and antisymmetric eigenstates of the double well are 1200 (400) and 300(100) respectively.

For the false-color images of Fig.3A, the average ICCD camera signals imaging the spatially resolved intensity of the cavity emission recorded within 200 pump pulses was used. To avoid drifts of the fringe pattern of the interferometer signal shown in Fig.4 from short-term fluctuations of the cavity cut-off caused by acoustic vibrations, part of the cavity emission was for the corresponding measurements split off to allow for simultaneous monitoring of the emission spectrum with a spectrometer. For data analysis, only images of the interferometer output recorded with the observed cavity cut-off being constant to within 0.02nm were used. The shown false-color image data are the average of 50 correspondingly post-selected ICCD camera signals, allowing for the averaging of sets of data with an effectively wavelength-stabilized cavity cutoff frequency.

Beamsplitter description

In the low-temperature limit, microcavity photons only populate the ground state $\psi_s$, with no population to the excited states. In an optical physics beamsplitter description an arbitary initial photon input state is - in this idealized case - thus converted to the unique state $\psi_s$ which defines a fully irreversible, non-unitary process. The phase space accumulation of the photonic degrees of freedom is made possible by the thermal coupling to the dye molecular reservoir, which allows for sympathetic cooling to the temperature of the dye solution. We point out that at finite temperature, besides a possible population of excited optical modes, the coupling to the photo-excitable dye molecules also results in number fluctuations of the photon gas, which can be tuned from canonical to grand canonical depending on the relative size of the dye reservoir (27).

Towards many-photon entangled states

Using the techniques outlined above, it is possible to structure the mirror surface in a highly flexible way. In the paraxial limit, the mirror spacing directly gives rise to a potential within the 2d Schrödinger equation which describes the quantum dynamics of cavity photons. Besides linear arrays for microcavities, one can also realize more complicated structure like rings of microcavities or junctions in various geometries. Also, one- and two-dimensional periodic lattice potentials can be realized (see also (12) and references therein).

Potentials for photonic systems have first been investigated in polariton systems in different experimental works, but a thermalization into the ground state of the structures has not yet been achieved (5). In a double-well system, Josephson oscillations of polaritons have been successfully observed (9).

In the presence of photon interactions, many-body quantum states can become the ground state of cavity photons in lattice potentials, see e.g. Refs. 23-25. A promising approach to realize sufficiently strong photon-photon interactions is to use second-order ($\chi^{(2)}$) nonlinear materials in a setup where interaction effects are enhanced by cavities with resonances at both the fundamental and the $\chi^{(2)}$-nonlinearly converted optical frequencies. Such a doubly resonant cavity setup can yield a strong effective Kerr ($\chi^{(3)}$) nonlinearity (22), as being required to generate a photon-photon interaction on the two-particle level. Here, photons can either first be doubled and then undergo a parametric oscillator process (as considered in (22)), or the reverse order can be applied. The former approach results in a Bose-Hubbard type photon interaction, which in the limit of one- and two-photon populations is also found for the second approach. Combining this approach with the realization of tunnel-coupled cavity states allows to create, for example, systems described by bosonic Hubbard models, where the coupling to a thermal bath may allow to cool into a low-temperature state. Depending on the setup, bosonic Hubbard models with or without extra polarization degrees of freedom (i.e., an effective spin index) can be realized. We consider the second case in the following. A simple, well studied ground state of such a system is, for example, a Mott state realized for integer filling (23-25).

A bosonic Mott insulator carries, however, almost no entanglement. To realize highly entangled multi-photon states it is better to consider Hubbard models with a fractional filling, e.g., a half-filled case. For example, it would be interesting to realize small rings (35) of Hubbard models. A ring shaped arrangement of tunnel-coupled microcavities (or any other geometric arrangement) can easily be realized by the described cavity mirror structuring technique. To give a specific example, the ground state for a bosonic quantum ring with N=4 sites at half filling in the limit of strong interactions is given by $\psi_s = \frac{1}{\sqrt{8}}(\sqrt{2}(|1010> + |0101>)$
$+ |1100> + |0110> + |0011> + |1001>)$,
which is a nonseparable, entangled state. The numbers here indicate the corresponding photon population in the individual sites n=1..N. Note that the bosonic Hubbard model can be mapped to a quantum XX-model (and to free fermions) in the strong coupling limit, see Refs. (36,37) for a discussion of the ground state wave functions in finite systems. Similar to the present experiment, also in the case that the ground state is a quantum manybody state we expect that a thermalized distribution of eigenenergies of the system eigenstates is obtained when the thermalization is faster than loss (21), i.e. when photons are absorbed by the optically active dye molecules faster than they leave the cavity through mirror transmission, coupling to unconfined optical modes, finite dye quantum efficiency, and non-radiative losses of the nonlinear medium.

Note that a photonic realization of highly entangled quantum state offers unique further advantages compared to, e.g., realizations by ultracold atoms. Nonlocal correlations and entanglement properties can be extracted using the photons emitted by the microcavities with established experimental procedures (6). Entangled optical N-particles states are of also of interest for a broad range of problems in quantum communication and computation (38,39).

**References:**

1. L.H. Greene, J. Thompson, J. Schmalian, *Rep. Prog. Phys.* **80**, 030401 (2017).

2. C. Gardiner, P. Zoller, *Quantum Noise* (Springer, Berlin, 2004).

3. P.W. Milonni, J.H. Eberly, *Lasers* (Wiley, New York, 1988).

4. T. Schwartz, G. Bartal., S. Fishman, M. Segev, *Nature* **446,** 52 (2007).

5. I. Carusotto, C. Ciuti, *Rev. Mod. Phys.* **85,** 299 (2013).

6. X.-C. Yau et al., Nat. Photonics **6**, 225 (2012).

7. J. Kasprzak, et al., *Nature* **443**, 409 (2006).

8. J. Klaers, J. Schmitt, F. Vewinger, M. Weitz, *Nature* **468**, 545 (2010).

9. M. Abbarchi et al., *Nature Phys*. **9**, 275 (2012).

10. P. Cristofolini et al. *Phys. Rev. Lett.* **110**, 186403 (2013).

11. M. Milicevic et al., Phys. Rev. Lett. **118**, 107403 (2017).

12. D. Dung, et al., *Nat. Photonics* **11**, 565 (2017).

13. H. Ohadi et al., Phys. Rev. X **6**, 031032 (2016).

14. N.G. Berloff et al., *Nat. Materials* **16**, 1120 (2017).

15. A. Mazurenko et al., *Nature* **545**, 39 (2017).

16. J. Marelic, R.A. Nyman, *Phys. Rev. A* **91**, 033813 (2015).

17. S. Greveling, K.L. Perrier, D. van Oosten, Phys. Rev. A **98**, 013810 (2018).

18. See supplementary materials.

19. J. Klaers, F. Vewinger, M. Weitz, *Nature Phys.* **6**, 512 (2010).

20. P. Kirton, J. Keeling, *Phys. Rev. Lett.* **111,** 100404 (2013).

21. J. Schmitt *et al. Phys. Rev. A* **92,** 011602 (2015).

22. A. Majumdar, D. Gerace, *Phys. Rev. B* **87**, 235319 (2013).

23. M.J. Hartmann, F.G. Brandao, M.B. Plenio, *Nature Phys.* **2**, 849 (2006).

24. A.D. Greentree, C. Tahan, J.H. Cole, L.C.L. Hollenberg, *Nature Phys.* **2**, 856 (2006).

25. D.G. Angelakis, M.F. Santos, S. Bose, *Phys. Rev. A* **76**, 031805 (2007).

26. A.W. De Leeuw, O. Onishchenko, R.A. Duine, H. T. C. Stoof, *Phys. Rev. A* **91**, 033609 (2015).

27. J. Schmitt, et al., *Phys. Rev. Lett.* **116**, 033604 (2016).

28. W. Verstraelen, M. Wouters, *Phys. Rev. A* **100**, 013804 (2019).

29. E. De Angelis, F. De Martini, P. Mataloni, *J. Opt. B* **2**, 149 (2000).

30. H. Yokoyama, S. D. Brorson, *J. Appl. Phys.* **66**, 4801 (1989).

31. J.R. Lakowicz, *Principles of Fluorescence Spectroscopy* (Kluwer Academic/Plenum Publishers, 1999).

32. V. Bagnato, D. Kleppner, *Phys. Rev. A* **44**, 7439 (1991).

33. W.J. Mullin, *J. Low Temp. Phys.* **106**, 615 (1997).

34. N. Tammuz et al., *Phys. Rev. Lett.* **106**, 230401 (2011).

35. K.M. O'Connor, W. K Wootters, Phys. Rev. A **63**, 052302 (2001).

36. E. Lieb, T. Schultz, D. Matthis, Ann. Phys. **16**, 407 (1961).

37. A. De Pasquale et al., Eur. Phys. J. Spec. Top. **160**, 127 (2008).

38. L. Amico, R. Fazio, A. Osterloh, V. Vedral, Rev. Mod. Phys. **80**, 517 (2008).

39. M. Epping, H. Kampermann, C. Macciavello, D. Bruß, New J. Phys. **19**, 093012 (2017).

**Figures:**

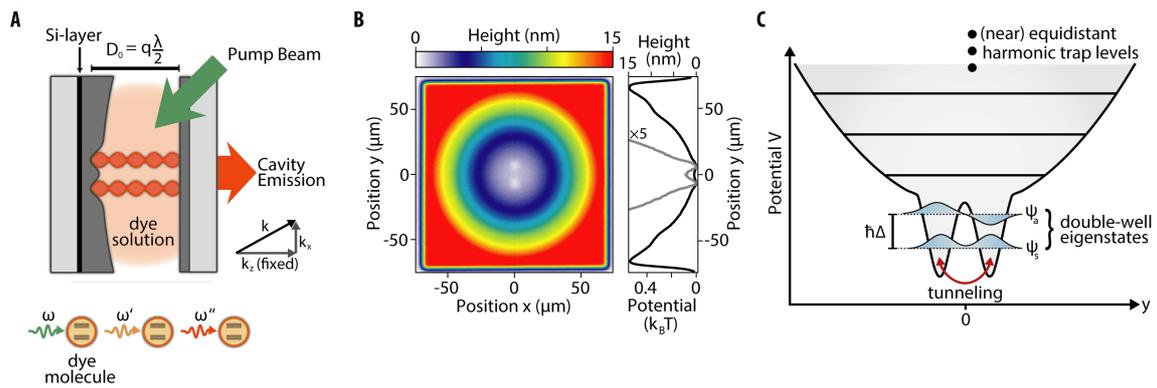

**Fig. 1. Experimental environment. (A)** Photons are trapped in a microresonator with a mirror spacing corresponding to an optical path of 5.5 wavelengths, where one mirror is laterally microstructured. The photons thermalize by repeated absorption re-emission processes (see bottom schematics) on dye molecules. **(B)** Height profile of the microstructured mirror surface (main plot) and profile of the corresponding expected trapping potential for cavity photons (right), realizing a double-well structure in the center superimposed by a harmonic trapping potential. **(C)** Schematic energy level structure, with the symmetric eigenstate of the double-well as the lowest energetic eigenstate.

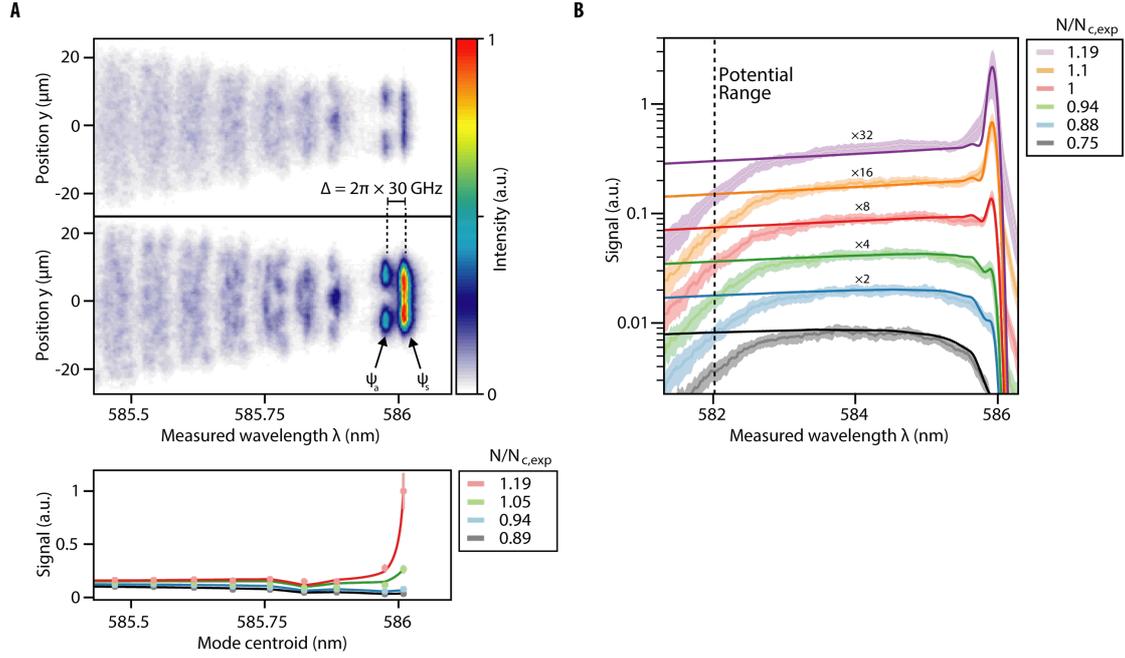

**Fig. 2. Spectrally resolved measurements.** (**A**) Measured wavelength versus transverse position along the axis of the double-well both below (top, $N/N_{c,exp} \cong 0.94$) and above (middle, $N/N_{c,exp} \cong 1.05$) the threshold to a Bose-Einstein condensate (18). The high resolution spectra cover the range of the first 9 lowest energetic (highest wavelength) modes. From right to left, the first two modes are the symmetric and antisymmetric modes of the double-well, followed by modes not confined in the wells which for the higher order modes are well described by harmonic oscillator modes with a trap frequency of $\Omega/2\pi \cong 63(2)$ GHz. The enhanced emission of the symmetric superposition observed in the latter image is attributed to Bose-Einstein condensation. The bottom panel gives spectra for different photon numbers. (**B**) Broadband, low-resolution spectra for different photon numbers (dots), along with theory for the dye microcavity temperature T=300K. The observed width of the BEC peak is dominated by spectrometer resolution. The spectral position of the rim of the potential well is indicated by the dashed line. For all measurements the critical photon number is $N_{c,exp} \cong 8000$.

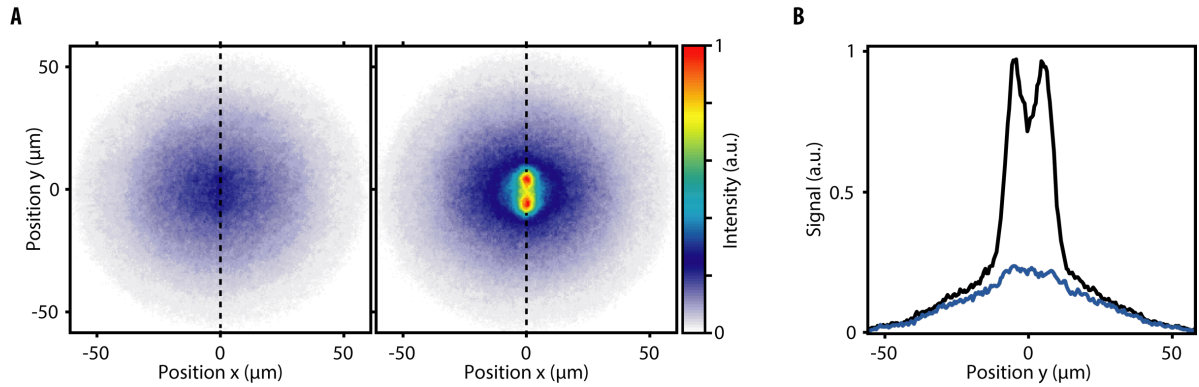

**Fig.3. Imaging the emission pattern.** (**A**) Spatial images of the cavity emission, both below (left) and above (right) the critical photon number (for $N/N_{c,exp} \cong 0.89$ and $1.19$ respectively). The double peak in the latter image on top of the thermal cloud is an in situ image of the split condensate. (**B**) Cuts along the direction of the double-well, indicated by the dotted lines in (A), for both the thermal cloud (blue) and the condensed phase (black) data.

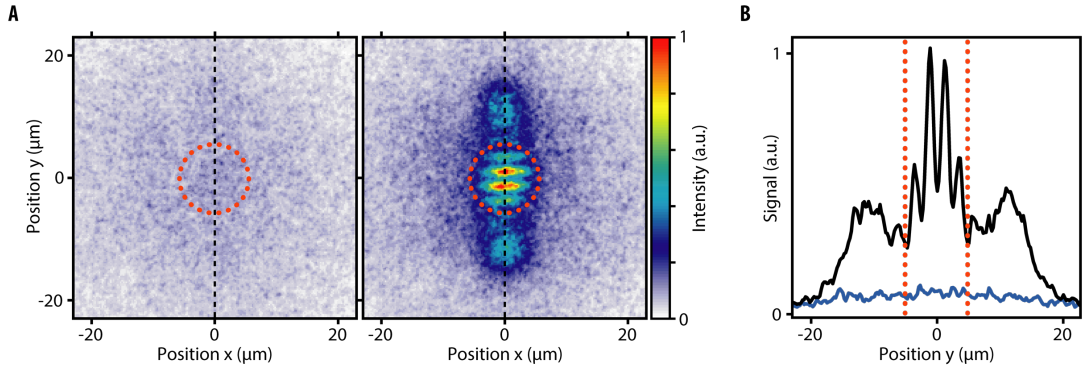

**Fig.4. Relative phase of the microsites emission. (A)** The emission from the two individual microsites is overlapped (indicated by the marked central region in the camera images) with a small tilt. The resulting interference pattern is shown for total photon number below (left) and above (right) the critical photon number, where in the condensed phase a stable interference signal is observed. The data are the average of 50 ICCD images recorded in 1μs measurement time each, corresponding to the length of the pump beam pulses (18). **(B)** Corresponding cuts along the double-well axis above (black) and below (blue) criticality (for $N/N_{c,exp} \cong 1.18$ and 0.88 respectively).

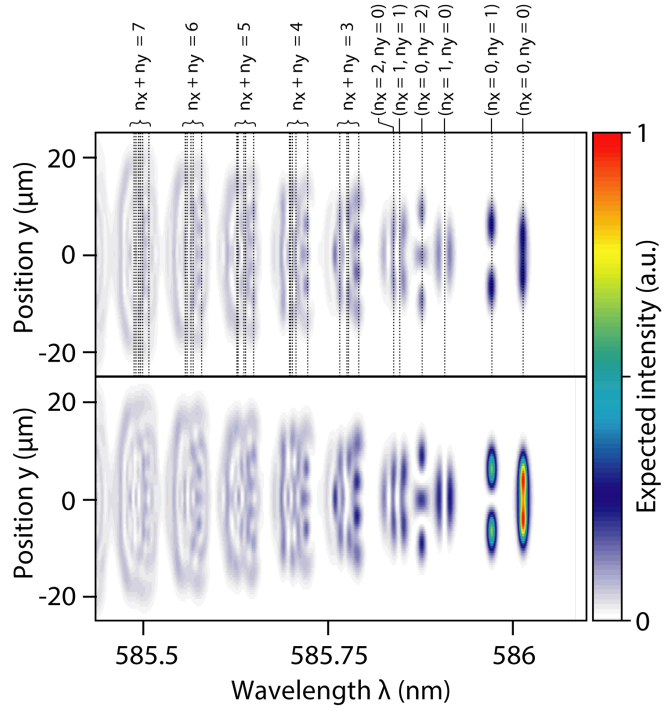

**Fig. S1.** Theoretically calculated photon intensity following spectral dispersion by the grating and demagnification along the x-axis, as derived from Eq. 4, for two values of the chemical potential, $\mu = -0.024\, k_B T$ (top) and $\mu = -0.006\, k_B T$ (bottom). The vertical scale gives the real position along the y-axis of the cavity plane and the horizontal axis encodes both the mode wavelength $\lambda_{n_x, n_y}$ and the position in x-direction, following $\lambda = \lambda_{n_x, n_y} + \kappa \cdot x$, as in our measurements. The results well describe the experimental data shown in Fig.2A of the main text. The parameter-free calculation is based on a numerical solution of the Schrödinger equation using the measured height profile of the microstructured cavity mirror and experimentally determined values of the chemical potential. In the upper panel the wavelength of each mode is shown as a dashed line, which are labeled by harmonic-oscillator quantum numbers, see text.